\documentclass[12pt]{article}
\usepackage{a4wide,graphicx}
\usepackage{cite}

\title{Searching for signatures around $1920$ MeV of a $N^*$ state of three hadron nature\\
\vspace{-3.5cm}
\hfill{{\tiny HISKP--TH--09/6, FZJ-IKP-TH--2009--6}}
\vspace{3.5cm}
}

\author{
A. Mart\'inez Torres$^1$\footnote{amartine@ific.uv.es}, K. P. Khemchandani$^2$\footnote{kanchan@teor.fis.uc.pt}, Ulf-G. Mei{\ss}ner$^{3,4}$\footnote{meissner@itkp.uni-bonn.de }, and E.~Oset$^1$\footnote{oset@ific.uv.es} \\
{\small{\it $^1$Departamento de F\'{\i}sica Te\'orica and IFIC,
Centro Mixto Universidad de Valencia-CSIC,}}\\
{\small{\it Institutos de
Investigaci\'on de Paterna, Aptdo. 22085, 46071 Valencia, Spain}}\\
{\small{\it $^2$ Centro de F\'isica Te\'orica, Departamento de F\'isica,}}\\
{\small{\it Universidade de Coimbra, P-3004-516 Coimbra, Portugal. 
}}\\
{\small{\it $^3$Helmholtz-Institut f\"ur Strahlen-und Kernphysik and Bethe Center for Theoretical Physics, }}\\
{\small{\it Universit\"at Bonn, D-53115 Bonn, Germany.}}\\
{\small{\it $^4$Institut f\"ur Kernphysik, Institute for Advanced Simulations
and J\"ulich Center for Hadron Physics}}\\
{\small{\it Forschungszentrum J\"ulich, D-52425 J\"ulich, Germany.}}\\
}

\date{\today}
\begin{document}
\maketitle

\begin{abstract} 
We provide a series of arguments which support the idea that the peak seen in
the $\gamma p \to K^+ \Lambda$ reaction around 1920 MeV should correspond to the
recently predicted state of $J^P=1/2^+$  as a bound state of 
$K \bar{K}N$ with  a mixture of  $a_0(980)N$ and $f_0(980)N$ components. 
At the same time we propose polarization experiments in that reaction 
as a further test of the
prediction, as well as a study of the total cross section for 
$\gamma p \to K^+ K^- p$ at energies close to threshold and of $d \sigma/d
M_{inv}$ for invariant masses close to the two kaon threshold.
\end{abstract}

\newpage

\section{Introduction}
The theoretical interest in three-hadron systems
 other than the traditional three-nucleon states is old. In 
 \cite{Harari:1964zz} there was a study of a possible system of
 $K \pi N$,  based only on symmetries. More recently, a possible
 $\bar{K}NN$ bound state has been the object of intense study
 \cite{shevchenko,dote,sato,akaishi}.
  However, lately there has been a qualitative step forward in
 this topic, made possible by combining elements of unitarized
 chiral perturbation theory $U\chi P T$
\cite{Kaiser:1995eg,Dobado:1996ps,Oller:1997ti,Oller:1998hw,ollerulf,granada,osaka}
 with Faddeev equations in coupled channels \cite{MartinezTorres:2007sr}.
In \cite{MartinezTorres:2007sr,MartinezTorres:2008is}
  systems of two mesons and one baryon with strangeness $S=-1$ 
 were studied, finding resonant states
which could be identified with the existing low-lying baryonic
$J^P=1/2^+$ resonances, two $\Lambda$ and four $\Sigma$ states. 
 Similarly, in the case of
the $S=0$ sector the $N^*(1710)$ appears neatly as a resonance of the $\pi \pi N$
system, as well as including the channels coupled to $\pi \pi N$ 
within SU(3) \cite{Khemchandani:2008rk}.  Further studies, including a 
realistic $\pi N$ amplitude beyond the $N^*(1535)$ region to which the chiral 
theories are limited, give rise to other $S=0$, $J^P=1/2^+$ states, more
precisely,  the $N^*(2100)$ and the  $\Delta(1910)$
\cite{MartinezTorres:2008kh}. 
Developments along the same direction produced a
resonant state of $\phi K \bar{K}$ \cite{MartinezTorres:2008gy} which could be 
identified with the X(2175) resonance reported by the BABAR collaboration
\cite{Aubert:2006bu,Aubert:2007ur} and later at BES \cite{:2007yt}.

  Independently, there have been similar ground-breaking studies based on 
variational methods \cite{jido,enyo}. In particular, in Ref.\cite{jido} a
bound state of $K \bar{K}N$ with  $I=1/2,J^P=1/2^+$
was found around 1910 MeV in the configuration $a_0(980)N$, suggesting a bound
state of the $a_0(980)$ and a nucleon. Since the $K \bar{K}$ system couples
to $\pi \pi$ and $\eta \pi$ to generate the $f_0(980)$ and $a_0(980)$ resonances,
a more complete coupled channels study using Faddeev equations was called for,
and this was done in 
\cite{MartinezTorres:2008kh} where it was concluded that a state appears indeed
around this energy, mostly made of $K \bar{K}N$ but in a mixture of  
$a_0(980)N$ and $f_0(980)N$.

  A $N^*$ state with these characteristics is not catalogued in the PDG
\cite{pdg}. However, there is a peak in the $\gamma p \to K^+
\Lambda$ reaction at around 1920 MeV, clearly visible in the integrated 
cross section and also at
all angles from forward to backward \cite{saphir,jefflab,mizuki}. The 
$\gamma p \to K^+ \Lambda$ reaction has been the object of intense theoretical
study \cite{Mart:1999ed,Usov:2005wy,Shklyar:2005xg,JuliaDiaz:2006is,bora} 
(see \cite{mart} for a recent review). With respect to  our present investigation,
the possible signal for a new resonance from the peak of the cross section
around 1920 MeV was already suggested in  \cite{Mart:1999ed}. However, no
spin and parity assignment were given, since there were several candidates 
in this region related to  the missing resonances of the quark models.  
Other theoretical studies do
not make use of this extra state, as in \cite{Usov:2005wy}, although in this
latter work only resonances with mass up to 1800 MeV were 
included\footnote{Work to include more resonances in the approach of \cite{Usov:2005wy}
is underway \cite{scholten}}. In a recent combined analysis of the 
$\gamma p \to K^+ \Lambda$  reaction with other reactions \cite{Nikonov:2007br}, 
a claim for a $N^*$ resonance around 1900 MeV was made, however, the 
resonance was assumed to have $J^P=3/2^+$.  Note in this respect that the 
PDG quotes in its latest
edition that there is no evidence for this resonance in the latest analysis of
the GWU group \cite{Arndt:2006bf}. The state around this energy found 
in  \cite{jido,MartinezTorres:2008kh} has instead  $J^P=1/2^+$ quantum numbers. 

As one can see, the situation concerning this state and its possible nature 
is far from settled. In the 
present work we collect several arguments to make a case in favor of the
state predicted in \cite{jido,MartinezTorres:2008kh} with $J^P=1/2^+$. 

\section{Comparison of the $\gamma p \to K^+ \Lambda$ and 
$\gamma p \to K^+ \Sigma^0$ reactions}

     A peak of moderate strength on top of a large background is clearly seen for the $\gamma p \to K^+ \Lambda$ reaction around $1920$ MeV at all angles (see Fig. 18 of \cite{jefflab}). One can induce qualitatively a width for this peak of about 100 MeV or less. On the other hand, if one looks at the $\gamma p \to K^+ \Sigma^0$ reaction, one finds that starting from threshold a big large and broad structure develops, also visible at all angles (see Fig. 19 of \cite{jefflab}). The width of this structure is about 200-300 MeV. One can argue qualitatively that the relatively narrow peak of the
$\gamma p \to K^+ \Lambda$ reaction around 1920 MeV on top of a large background has nothing to do with the broad structure of $\gamma p \to K^+ \Sigma^0$ around 1900 MeV. A more quantitative argument can be provided by recalling that in \cite{lee} the broad structure of the 
$\gamma p \to K^+ \Sigma^0$ is associated to two broad $\Delta$ resonances in their model, which obviously can not produce any peak in the   $\gamma p \to K^+ \Lambda$ reaction, which filters isospin $1/2$ in the final state. Certainly, part of the structure is background, which is already obtained in chiral unitary theories at the low energies of the reaction  \cite{bora}.
   
\begin{figure}[t]
\centering
\includegraphics[width=\textwidth]{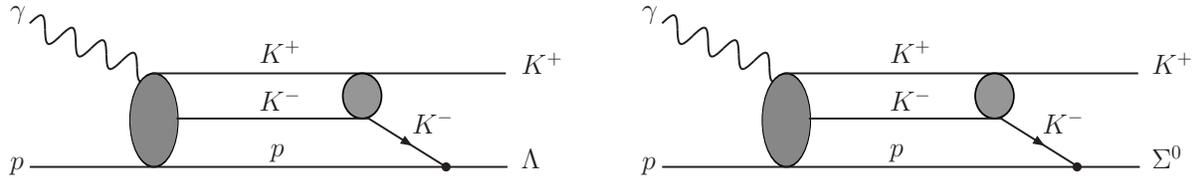}
\caption{Diagrams depicting the $\gamma p\rightarrow K^+\Lambda$, $\gamma p\rightarrow K^+\Sigma^0$ processes through the $1/2^+$ $N^*$ $K^+K^-N$ resonance of \cite{MartinezTorres:2008kh,jido}.}\label{gpKKL}
\end{figure}

    We thus adopt the position that the peak in the  $\gamma p \to K^+ \Lambda$ reaction is a genuine isospin $1/2$ contribution which does not show up in the $\gamma p \to K^+ \Sigma^0$ reaction. This feature would find a natural interpretation in the
picture of the state proposed in \cite{jido,MartinezTorres:2008kh}. Indeed, in
those works the state at 1920 MeV is a $K \bar{K} N $ system in relative 
s-waves for all pairs.  The $\gamma p \to K^+ \Lambda$ and 
$\gamma p \to K^+ \Sigma^0$ reactions proceeding through the excitation of 
this resonance are depicted
in Fig.~\ref{gpKKL}. The two reactions are identical in this picture,
the only difference being the Yukawa coupling of the $K^-$ to the proton to
generate either a $\Lambda$ or a $\Sigma^0$.

The Yukawa couplings in SU(3) are well known and given in terms of the F and D
coefficients  \cite{Bernard:1995dp}, with $D+F=1.26$ and $D-F= 0.33$ 
\cite{Close:1993mv,Borasoy:2007ku}.  The particular couplings for 
$K^- p\rightarrow \Lambda$ and $K^-p\rightarrow\Sigma^0$are e.g. given in~\cite{phimed}.

\begin{eqnarray}
V_{K^- p\rightarrow\Lambda}&=&-\frac{2}{\sqrt{3}}\frac{D+F}{2f}+\frac{1}{\sqrt{3}}\frac{D-F}{2f}\nonumber\\
\\
V_{K^- p\rightarrow\Sigma^0}&=&\frac{D-F}{2f}\nonumber
\end{eqnarray}
with $f$ the pion decay constant. Hence, the couplings are proportional to $-1.26$
and $0.33$ for the $K^- p \to \Lambda$ and $K^- p \to \Sigma^0$ vertices,
respectively. Therefore, it is clear that in this picture the signal of the
resonance in the $\gamma p \to K^+ \Lambda$ reaction is far larger than in the
$\gamma p \to K^+ \Sigma^0$  one, by as much as an order of magnitude in the 
case that
the resonance and background contributions  sum incoherently. The $3/2^+$ resonance 
used in the analysis of the $\gamma p \to K^+ \Lambda$ reaction in \cite{Nikonov:2007br} 
is also used for the $\gamma p \to K^+ \Sigma^0$  reaction in that work and also give a smaller contribution in this latter case. In the picture
of \cite{jido,MartinezTorres:2008kh} the $1/2^+$ resonance also appears in 
the $\gamma p \to K^+ \Sigma^0$ reaction but with a smaller intensity than in the 
$\gamma p \to K^+ \Lambda$ one, as we have mentioned.

\section{Pion induced reactions}

\begin{figure}[h!]
\centering
\includegraphics[width=\textwidth]{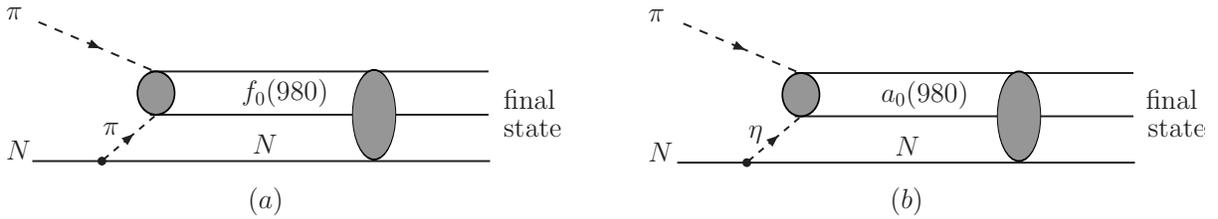}
\caption{Diagrams for $\pi$ induced reactions exciting the $1/2^+$ 
resonance of~\cite{MartinezTorres:2008kh,jido}.}\label{ppN}
\end{figure}

  The fact that a $N^*$ resonance around 1900 MeV is not reported in the
PDG tables finds a natural interpretation in the picture of 
\cite{jido,MartinezTorres:2008kh}. Indeed, since most of the information about
resonances is obtained from  $\pi N$ reactions, it is easy to
understand why this resonance did not appear in these reactions. Once again, in
the picture of \cite{jido,MartinezTorres:2008kh}, the pion induced reactions
going through the resonance would proceed in a direct way as shown in Fig.~\ref{ppN}a.

As we can see from that figure, the fact that the $f_0(980)$ has a small
coupling to  $\pi \pi$ \cite{Oller:1997ti}, reflected in its small decay width,
necessarily weakens the strength of the pion induced reactions producing the
resonance, compared to other processes. Since the wave function of the state
has also an $a_0(980)N$ component, in this case the mechanism would be the one
of Fig. \ref{ppN}b, since the $a_0(980)N$ has also a small coupling to $\eta \pi$. Thus,
the smallness of this coupling, and also the small $\eta NN $ coupling
(see e.g. Ref.~\cite{Meissner:1988iv}), make again the mechanism of production very weak.

\begin{figure}[h!]
\centering
\includegraphics[width=\textwidth]{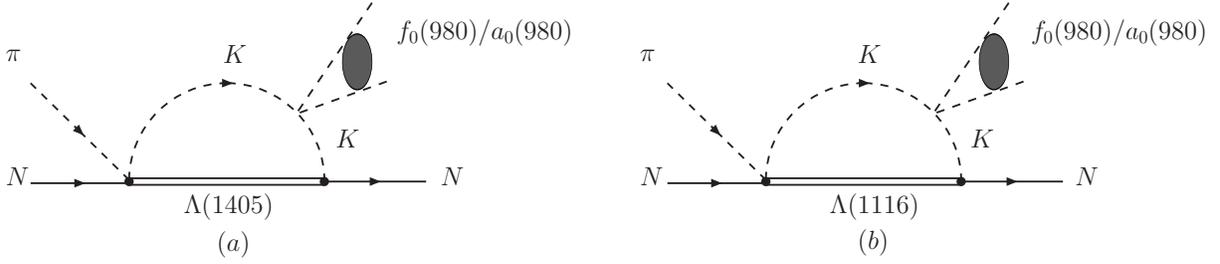}
\caption{Possible indirect mechanism for the $\pi N\to N f_{0}(980)/a_{0}(980)$.}\label{KL}
\end{figure}

It is possible to devise some indirect method to create the $Nf_{0}(980)/a_{0}(980)$ intermediate states. One can devise the mechanisms of Fig. \ref{KL}.
The diagram of Fig. \ref{KL}a involves the $\pi N\to K\Lambda(1405)$ transition while the one of Fig. \ref{KL}b involves the $\pi N\to K\Lambda$ one. The strength of the former amplitude is much weaker than that of the second one as one can induce from experimental cross sections \cite{thomas,prl} and the associated amplitudes \cite{hyodo,inoue}. However, the second term is penalized by the $p$-wave coupling $KN\Lambda$ in the loop, where the two other vertices are in $s$-wave. As a general rule, loops are reduced with respect to tree level amplitudes, but in the present case the structure of these loops, with a meson-meson $\to$ meson-meson vertex, makes the contribution very small as a direct evaluation of the diagrams shows \cite{vmagas}.

\begin{figure}[h!]
\centering
\includegraphics[width=\textwidth]{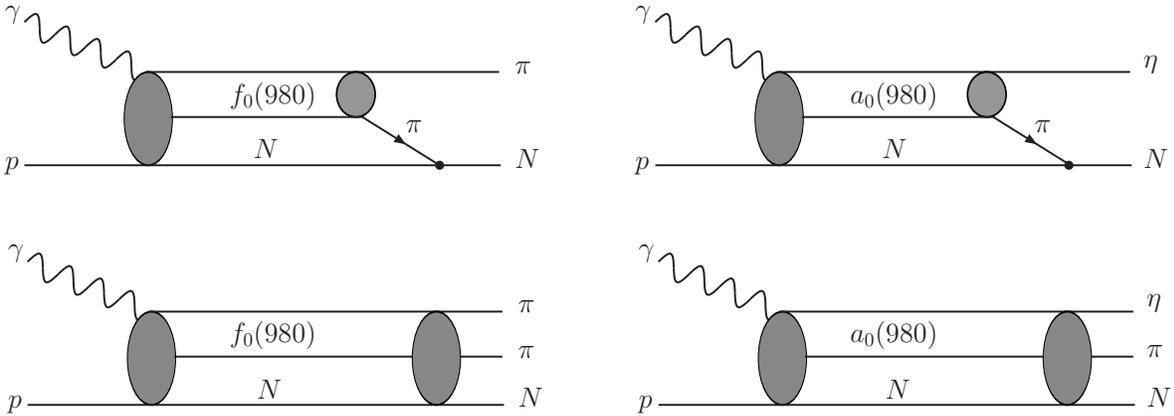}
\caption{Diagrammatic representation of the $\gamma$ induced reactions through the $1/2^+$ $N^*$ resonance of \cite{MartinezTorres:2008kh,jido} with $\pi N$, $\eta N$, $\pi\pi N$ or $\eta\pi N$ in the final state.}\label{gpppN}
\end{figure}

  In this respect it is also very illustrative to see that in \cite{Nikonov:2007br}
a large set of reactions was analyzed, and in Table 1 of the paper it was shown
that the resonance around 1900 MeV had a weight bigger than 1 \% only in the 
$\gamma p \to K^+ \Lambda$ and  $\gamma p \to K^+ \Sigma^0$ reactions. This 
means that the
weight of this resonance  in the $\gamma p \to \pi N$, $\gamma p \to \eta N$,
$\gamma p \to \pi \pi N$, $\gamma p \to \pi \eta N$ and $\pi N \to \pi \pi N$,
analysed there,  is
negligible. This would again find a natural explanation along the lines
discussed above if one looks at the mechanisms for these reactions depicted in
Fig. \ref{gpppN} which are all suppressed, since they always involve the
coupling of the $f_{0}(980)$ to $\pi\pi$ or the $a_{0}(980)$ to $\pi\eta$.

\section{Angular distributions}

\begin{figure}[t]
\centering
\hspace{-05cm}\includegraphics[scale=0.8]{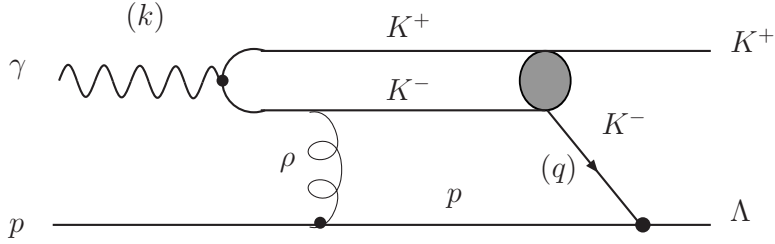}
\caption{Schematic amplitude for $\gamma p\rightarrow K^+\Lambda$ exciting the $1/2^+$ $N^*$ resonance of \cite{MartinezTorres:2008kh,jido}.}\label{gpipKKbarL}
\end{figure}

 In this section we make a different argumentation in favor of the
 interpretation of the state of \cite{jido,MartinezTorres:2008kh}. We study the
angular dependence of the $\gamma p \to K^+ \Lambda$ proceeding through the
resonance excitation. In order to get the basic structure of the amplitude
with the quantum numbers of the resonance we take a typical mechanism
compatible with the nature of the resonance as having $ K \bar{K}$ in s-wave and
in relative s-wave with the nucleon. This is shown in Fig. \ref{gpipKKbarL}

The structure of the amplitude, close to $K^+K^- p$ threshold, is given by
\begin{equation}
t^{(1/2)}\propto\vec{\sigma}\vec{q}\,
\Bigg(\frac{\vec{\sigma}\times\vec{k}}{2M_{N}}\Bigg)\,\vec{\epsilon}\label{th}
\end{equation}
where $q$ ($\vec{q}=-\vec{q}_{K^+}$) and $k$ are depicted in the figure, $M_{N}$ is the nucleon mass an
$\vec{\epsilon}$ is the photon polarization vector. The amplitude can be rewritten as
\begin{equation}
t^{(1/2)}\propto\vec{\epsilon}\,(\vec{q}\times\vec{k})+i\vec{\epsilon}\,\vec{q}\,\vec{\sigma}\,\vec{k}-i\vec{\epsilon}\,\vec{\sigma}\,\vec{k}\,\vec{q}
\end{equation}
Summing the modulus squared of the amplitude over initial and final 
polarizations of the nucleons and the photons one obtains
\begin{equation}
\overline{\sum}\sum |t^{(1/2)}|^2\propto 2\vec{q}^{\,2}\,\vec{k}^{\,2}
\end{equation}
and we see that there is no angular dependence.

\begin{figure}[h!]
\centering
\hspace{-05cm}\includegraphics[scale=0.8]{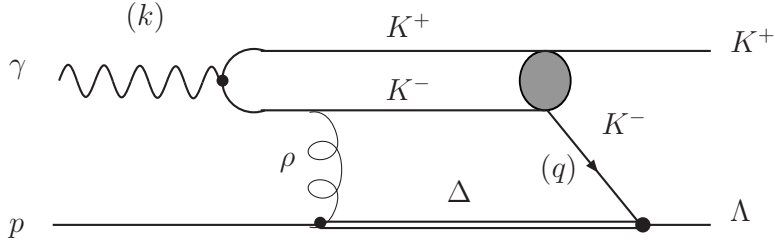}
\caption{Schematic amplitude for $\gamma p\rightarrow K^+\Lambda$ exciting a 
$3/2^+$ $\Delta$ intermediate state.}\label{gpiKKbarL}
\end{figure}

Next we assume that we have a $J^P=3/2^+$ state and we show in Fig.~\ref{gpiKKbarL} 
the equivalent diagram to that in Fig. \ref{gpipKKbarL}, but assuming a $J^P=3/2^+$ baryonic 
intermediate state coupled to $K \bar{K}$. The amplitude has now the structure
\begin{equation}
t^{(3/2)}\propto\vec{S}\vec{q}\,\Bigg(\frac{\vec{S}^\dagger\times 
\vec{k}}{2M_{N}}\Bigg)\, \vec{\epsilon}
\end{equation}
with $\vec{S}$ the spin transition operator from spin 3/2 to spin 1/2. We can
rewrite the amplitude taking into account the relationship
\begin{equation}
\sum_{M_{S}}S_{i}|M_{S}\rangle\langle M_{S}|S^\dagger_{j}=\frac{2}{3}\delta_{ij}-\frac{i}{3}\epsilon_{ijk}\sigma_{k}
\end{equation}
and find
\begin{equation}
t^{(3/2)}\propto\frac{2}{3}\vec{\epsilon}\,(\vec{q}\times\vec{k})-\frac{i}{3}\vec{\epsilon}\,\vec{q}\,\vec{\sigma}\vec{k}+\frac{i}{3}\vec{\epsilon}\,\vec{\sigma}\,\vec{k}\,\vec{q}
\end{equation}

Upon sum of the modulus square of the amplitudes over initial and final spins we
find now

\begin{equation}
\overline{\sum}\sum |t^{(3/2)}|^2\propto\frac{1}{9}\vec{k}^{\,2}\,\vec{q}^{\,2}(3\,sin^2\theta+2)
\end{equation}
We see that now we have a strong angular dependence and the biggest strength
is expected  for $\theta$ = 90 degrees. 

\begin{figure}[h!]
\centering
\includegraphics[width=\textwidth]{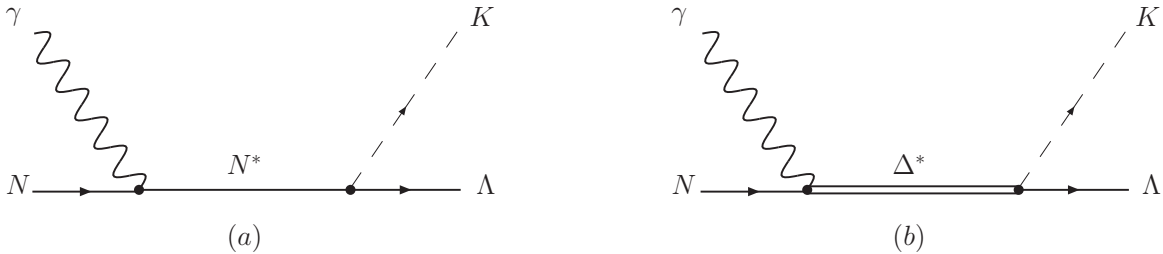}
\caption{Diagrams for the process $\gamma N\to N^*\to K\Lambda$ (a) and $\gamma N\to \Delta^* \to K\Lambda$ (b).}\label{new}
\end{figure}

Although we have extracted the angular dependence from the particular model of Figs. \ref{gpipKKbarL}, \ref{gpiKKbarL},
the results are general for $\gamma N\to R\to K\Lambda$ with $R$ a $1/2^+$ or a $3/2^+$ resonance, as one would get from the tree level amplitudes of Fig. \ref{new} using the standard $\gamma NN$, $P NN$, $\gamma NR$, $PNR$ vertices, with $P$ standing for a pseudoscalar meson \cite{carrasco}.

   Next we would like to recall what happens experimentally. As one can see in
Ref.~\cite{jefflab}, the signal around 1920 MeV  is present at all angles and
one finds roughly a contribution of the peak around $1920$ MeV over a smooth 
background of the order of $0.5\,\mu b$ for ${d\sigma}/{dcos\theta}$ at all
angles. This would disfavor the association
of the peak to a $3/2^+$ resonance, since in this case  at 90 degrees
one would expect the maximum signal of the resonance, with a strength about 
$5/2$ the size of the one at forward or backward angles. The argument assumes that
signal and background will sum incoherently in the reaction, which does not
have to be necessarily correct, but one does not expect much coherence either in
view of the many mechanisms contributing to the background in the theoretical
models. 

    We have also given some thought to the possible use of the asymmetries already measured for this reaction \cite{mizuki,zegers}. It is easy to see that the asymmetry, $\Sigma$, for the amplitude of our $1/2^+$ state given by Eq.$\,$(\ref{th}) is $\Sigma=0$. With this value of $\Sigma$, and assuming the contribution of the peak of the state over the background of about $20\%$, we find that adding the contribution of the $1/2^+$ signal $\Sigma$ is reduced by about $20\%$. Considering that the values measured for $\Sigma$ in \cite{mizuki,zegers} have less precision than the cross sections, $\Sigma \simeq 0.25\pm 0.15$, the changes induced in $\Sigma$ by the $1/2^+$ signal are of no much help.
    
\section{Test with polarization experiments}

In case the $J^P=1/2^+$ assignment was correct, an easy test can be carried out to rule out the $3/2^+$ state. 
The experiment consists in performing the 
$\gamma p \to K^+ \Lambda$ reaction with a
circularly polarized photon with helicity~1, thus $S_z=1$ with the $z$-axis
defined along the photon direction, together with a polarized proton of the
target with $S_z=1/2$ along the same direction. With this set up, the total spin
has $S^{tot}_z=3/2$. Since $L_z$ is zero with that choice of the $z$ direction,
then $J^{tot}_z=3/2$ and $J$ must be equal or bigger than $3/2$. Should the
resonant state be $J^P=1/2^+$, the peak signal would disappear for this polarization
selection, while it would remain if the resonance was a $J^P=3/2^+$ state. 
Thus, the disappearance of the signal with this polarization set up would rule out the 
$J^P=3/2^+$ assignment.

  Such type of polarization set ups have been done and are common in facilities
like ELSA at Bonn, MAMI B at Mainz or CEBAF at Jefferson Lab, where spin-3/2
and 1/2 cross sections, which play a crucial role
in the GDH sum rule, see e.g. Ref.~ \cite{Drechsel:1995az}, were measured in the
two-pion photoproduction \cite{Ahrens:2001qt,Ahrens:2007zzj} reaction.
The theoretical analysis of \cite{Nacher:2001yr} shows indeed that the
separation of the amplitudes in the spin channels provides information on the
resonances excited in the reaction.

\section{Analysis of the reaction $\gamma p \to K^+ K^- p$ close to threshold}

\begin{figure}[h!]
\centering
\hspace{-03cm}\includegraphics[scale=0.6]{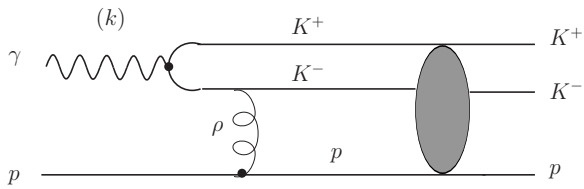}
\caption{Schematic amplitude for $\gamma p\rightarrow K^+K^-p$ exciting the $1/2^+$ $N^*$ resonance of \cite{MartinezTorres:2008kh,jido}.}\label{gpiKKbarp}
\end{figure}

 An ideal test of the nature of the resonance predicted around 1920 MeV is the
study of the process $\gamma p \to K^+ K^- p$ close to threshold. This reaction has
received relatively good experimental attention
\cite{Barber:1980zv,Barth:2003bq,Kubarovsky:2006ns,Nakano:2007zz}, stimulated 
recently by the search of a possible pentaquark state of strangeness S=1. 
Theoretically it has also been studied in 
\cite{Roberts:2004rh,Oh:2004wp,Roca:2004wt,Roca:2006sz,Sibirtsev:2005ns},
also motivated by the search of the pentaquark of strangeness $S=1$, or the nature of the
$\Lambda(1520)$  and other resonances. Yet, the emphasis here is different and
the measurements required are very close to threshold, in the region below the
$\phi$ production to avoid unnecessary complications in the analysis.
Indeed, as we
mentioned, the resonance has all the $K^+ K^- p$ components in s-wave and the energy is
just a bit below the threshold of the reaction. The effect of this resonance should be
seen as an accumulation of strength in the cross section close to threshold
compared with phase space. Such effects are common in many reactions
\cite{Oset:2001ep,Hanhart:2007yq,Dzyuba:2008pg}. In  order to see the effects
expected for this case we again draw in Fig.~\ref{gpiKKbarp} a typical diagram which would 
contribute to the $\gamma p \to K^+ K^- p$ close to threshold through the 
intermediate resonance excitation.

By looking at Fig.~\ref{gpiKKbarp}, in the shaded blob of the figure one is
assuming all interactions of $K^+K^-p\rightarrow K^+K^-p$, which one
encounters in a full  Faddeev  calculation. Hence, the amplitude for 
$\gamma p\rightarrow K^+K^-p$ can be written as
\begin{equation}
t_{prod}\propto T_{K^+K^- p\rightarrow K^+K^- p}\label{tprop}
\end{equation}
The $T_{K^+ K^- p \to K^+ K^- p}$ amplitude was evaluated using Faddeev
equations in coupled channels \cite{MartinezTorres:2008kh}. We are assuming
that this term, which exhibits the  peak due to the $1/2^+$ $N^*$ resonance, 
dominates over a possible background term close to threshold (for instance, a
tree level diagram like the one of Fig. \ref{gpiKKbarp} omitting the
interaction of the particles symbolized by the shaded blob or diagrams with
two-particle final-state interactions). The scheme of the latter work is very rewarding
for  experimental tests. Indeed, what one evaluates there is the $T$-matrix as a
function of two variables. These are $\sqrt{s}$, the total energy, and $\sqrt{s_{23}}$,
the invariant mass of the subsystem of two particles that one expects to be
highly correlated. This is the case here, where the $K^+ K^-$ is correlated to
the $a_0(980)$ and $f_0(980)$ resonances below the $K^+ K^-$ threshold, hence 
$\sqrt{s_{23}}$ is taken for the $K^+ K^-$ pair. 
Since the $T$ amplitude of \cite{MartinezTorres:2008kh} does not have any angular
dependence, and for a fixed total energy only depends on $\sqrt{s_{23}}$, the
differential cross section for the $\gamma p \to K^+ K^- p$ reaction is readily
found and we have

\begin{eqnarray}
\frac{d\sigma}{d M_{inv}}&=&C\frac{1}{s-M_{N}^2}
\frac{1}{\sqrt{s}}p\,\tilde{q}|T_{K^+K^- p\rightarrow K^+K^- p}|^2\label{dsigma}\\
p&=&\frac{\lambda^{1/2}(s,M^2_{inv},M_{N}^2)}{2\sqrt{s}},\quad\quad\tilde{q}=\frac{\lambda^{1/2}(M^2_{inv},m^2_{K},m^2_{K})}{2M_{inv}}
\end{eqnarray}
where $C$ is a constant and we have written $M_{inv}$ for the invariant mass of the two-kaon system, 
the $\sqrt{s_{23}}$ variable in Ref.~\cite{MartinezTorres:2008kh}.

\begin{figure}[t]
\centering
\includegraphics[scale=0.5,angle=0]{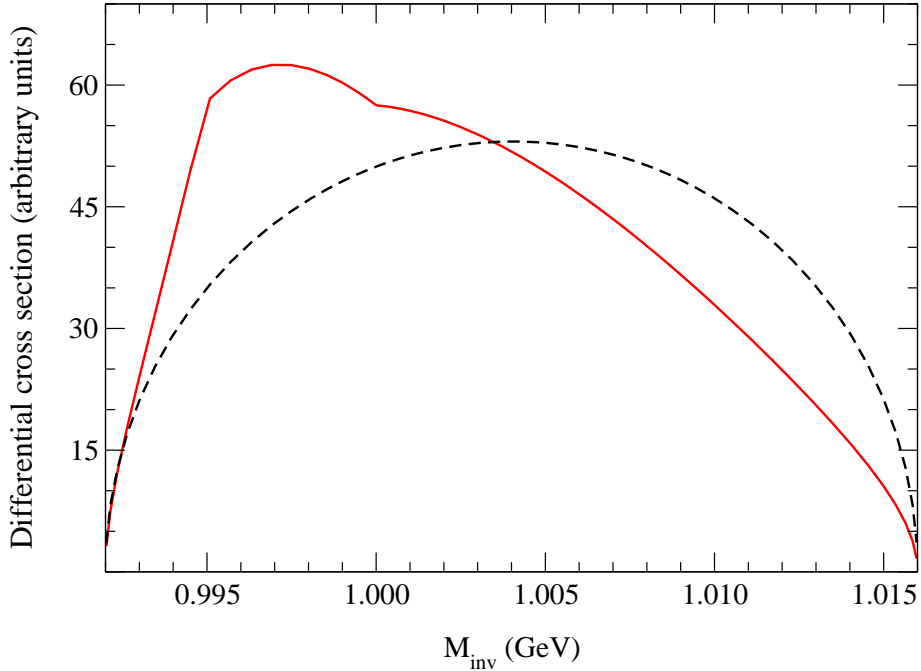}
\caption{${d\sigma}/{dM_{inv}}$ with phase space (dashed line) and using the 
$K^+K^- p\rightarrow K^+K^-p$ amplitude of \cite{MartinezTorres:2008kh} 
(solid line), with $M_{inv}$ the invariant mass for the $K^+K^-$ system. The curves have been normalized to have the same integrated 
cross section.}\label{dcross}
\end{figure}

\begin{figure}[t]
\centering
\includegraphics[scale=0.5,angle=0]{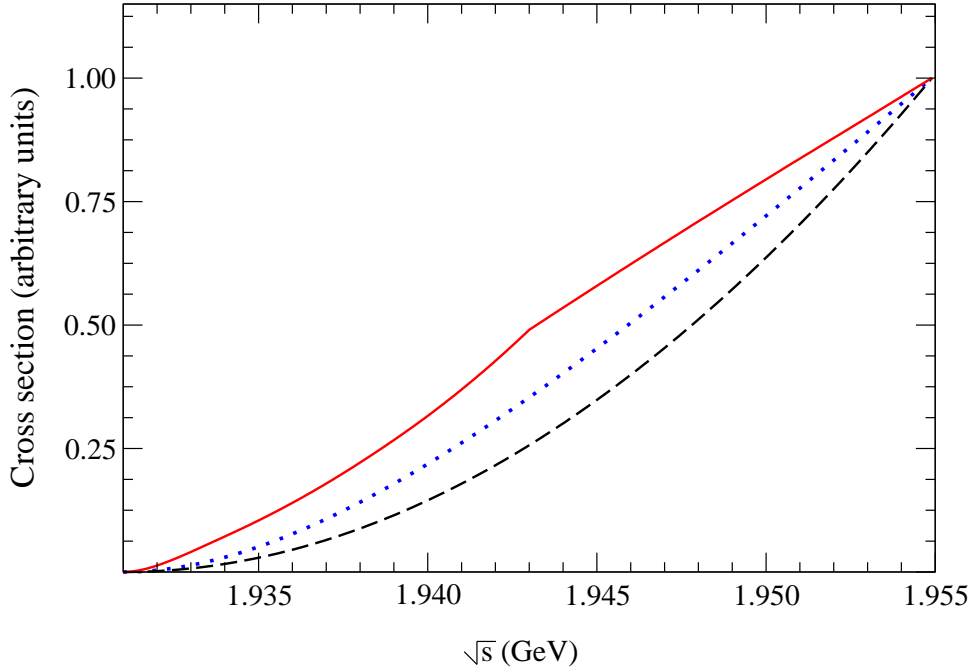}
\caption{Integrated cross sections. Dashed line: phase space. Dotted line: 
using the amplitude of Eq.$\,$(\ref{Tf0}) accounting for $K^+K^-$ final-state 
interaction. Solid line: results obtained using the $K^+K^- p\rightarrow
K^+K^- p$ amplitude of~\cite{MartinezTorres:2008kh}. 
The curves have been normalized to unity at $1955$ MeV.}\label{cross}
\end{figure}

In order to show the effects of the resonance below threshold and the
correlations of $M_{inv}$ we show two plots in Figs. \ref{dcross} and \ref{cross}.

In Fig.~\ref{dcross} we take a fixed energy, just below  $\phi N$
production  with
$\sqrt{s}=1955$ MeV and plot $d \sigma/d M_{inv}$ as a function of  $M_{inv}$
for phase space (dashed line) and for Eq.$\,$(\ref{dsigma}) (solid line).   
As we can see, there is a big asymmetry of
the mass distribution with respect to phase space, with a clear accumulation of
strength close to  $M_{inv}= 2 m_K$ as a consequence of the presence of the
$f_0(980)$ or $a_0(980)$ below threshold. The results have been normalized 
to the same integrated cross section.

In Fig.~\ref{cross}, we have instead represented the integrated cross section of 
$\gamma p \to K^+ K^- p$ as a function of the energy from threshold up to 
$\sqrt{s}=1955$ MeV evaluated with Eq.$\,$(\ref{dsigma}) (solid line) and we 
compare the results with phase space (dashed line). The cross section have
been normalized 
to one at $\sqrt{s}=1955$ MeV for comparison. What we observe here is that 
the cross section is also more pronounced
at lower energies as a consequence of the presence of the three particle 
resonance below threshold. 

We should also take into account that the shape of ${d\sigma}/{d M_{inv}}$ 
in Fig.~\ref{dcross} should be expected from final-state interactions of the
$K^+ K^-$ pair close to threshold, even if the $N^*(1920)$ resonance were not 
present \cite{Oset:2001ep,Hanhart:2007yq,Dzyuba:2008pg}. This means that
instead of phase space we should already consider a $T$ matrix element
accounting for the $K^+K^-$ final-state interaction incorporating the pole 
of the $f_{0}(980)$ or $a_{0}(980)$. Thus we perform the evaluation of the
cross section with an empirical amplitude
\begin{equation}
t_{prod}\propto\frac{1}{M_{inv}^2-M_{f_{0}}^2+i\Gamma_{f_{0}}M_{f_{0}}}\label{Tf0}
\end{equation}
with $M_{f_{0}}=980$~MeV and $\Gamma_{f_{0}}=30$~MeV. We note that using this 
amplitude we obtain a distribution ${d\sigma}/{dM_{inv}}$ very similar to the 
one obtained in Fig.~\ref{dcross} using the amplitude
of~\cite{MartinezTorres:2008kh}. 
The results for the integrated cross section with the amplitude of
Eq.$\,$(\ref{Tf0}) can be seen in the dotted line in Fig.~\ref{cross}. We can
still see deviations from the new curve incorporating the $K^+K^-$ final-state 
interaction with respect to the curve accounting for the $N^*(1920)$ resonance 
in addition (solid curve in Fig.~\ref{cross}).

\begin{figure}[t]
\centering
\includegraphics[scale=0.5,angle=0]{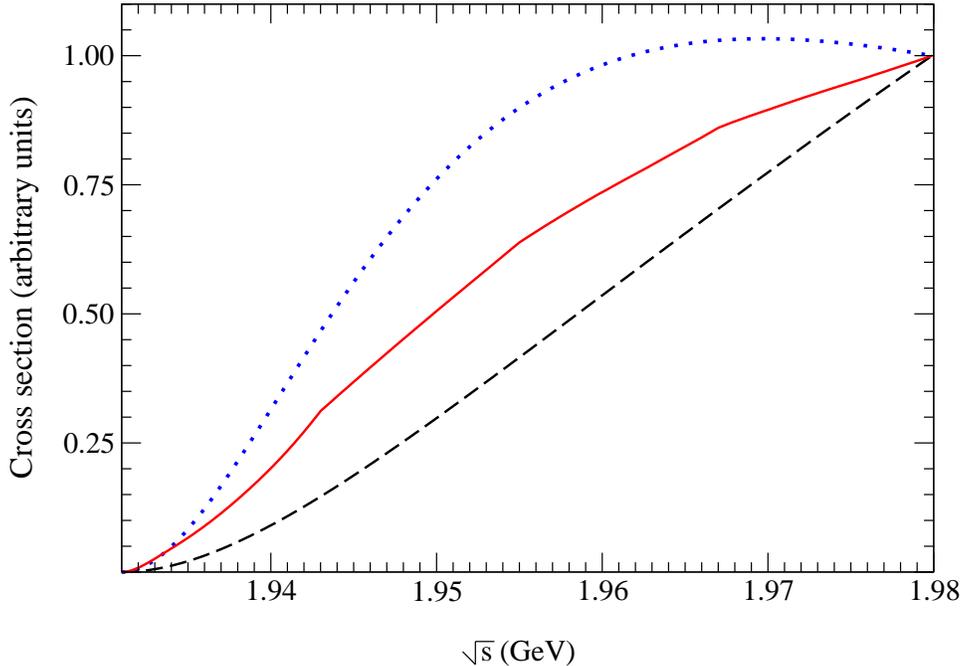}
\caption{Integrated cross sections. Dashed line: using the amplitude of 
Eq.$\,$~(\ref{Tf0}) which accounts for the $K^+K^-$ final-state interaction. 
Solid line: results using the $K^+K^-p$ amplitude of
\cite{MartinezTorres:2008kh}. 
Dotted line: results using the empirical amplitude of Eq.$\,$~(\ref{Tf0BW}) 
accounting for the $K^+K^-$ final-state interaction plus a resonant pole 
at $1924$~MeV with a width of $60$~MeV.}\label{cross3}
\end{figure}

In the calculations of  Ref.~\cite{MartinezTorres:2008kh}
the width obtained, of around $20$ MeV, should be smaller than the one of the 
real state because one only has three-body states and the decay into two
particles is not included. As discussed in \cite{MartinezTorres:2008kh},  even
if the  building blocks are three particles, one can
obtain a larger width for the decay into two-body systems because there is 
more phase space for it.  Because of that, in order to estimate possible
uncertainties from this deficiency, we show what we would observe if we just had
a resonance below threshold with a typical Breit-Wigner amplitude with 
mass $M_{R}=1924$ MeV and width of about 60 MeV, 
\begin{equation}
t_{prod}\propto \frac{1}{\sqrt{s}-M_{R}+i\frac{\Gamma}{2}}\frac{1}{M_{inv}^2-M_{f_{0}}^2+i\Gamma_{f_{0}}M_{f_{0}}}\label{Tf0BW}
\end{equation}
The problem in this case is that the span of energies chosen in
Fig.~\ref{cross} from threshold to $1955$~MeV is only of about $25$ MeV,
smaller than the width, such that one can not  see the resonance structure in
such a narrow range any more. Going to higher energies has the problem that we 
should face the $\phi$ production which has a large cross section. A way out
of this problem could be found by eliminating from the experimental cross
section the very narrow peak for the $\phi$ production. Assuming then that
there is no contribution from $\phi$ production we compare the result obtained 
with the amplitude of Eq.$\,$(\ref{Tf0}) and the one with
Eq.$\,$(\ref{Tf0BW}) which accounts also for the 
$N^*(1920)$ resonant pole. We show the results in Fig.~\ref{cross3}, where the
cross sections have been normalized to one at $\sqrt{s}=1980$ MeV.

We observe that the effect of the $N^*(1920)$ resonance is clearly visible, 
with an enhanced cross section at lower energies compared with the curve that 
has only the $K^+K^-$ final state. We also include in the figure the results
obtained using the amplitude of Eq.$\,$(\ref{tprop}) (solid line). The region between the two upper curves can indicate the theoretical 
uncertainties but we can see that, in any case, a clear enhancement due to 
the $N^*$ resonance is seen.

 The effects observed in this calculation call for an experimental test which
 could verify, or eventually refute  the findings obtained above based on the
 nature of the predicted $1/2^+$ state around 1920 MeV.  This analysis is
 underway at Spring8/Osaka \cite{nakano}.

\section{Conclusions}
The independent prediction, also using different methods, of a $1/2^+$ state 
around 1920 MeV as a bound state of $K \bar{K} N $, which appears in the 
$a_0(980)N$ and $f_0(980)N$ configurations, stimulated us to suggest that this
state may have already been seen in the peak observed in the 
$\gamma p \to K^+ \Lambda$ reaction around this energy. Based on the structure
found in \cite{jido,MartinezTorres:2008kh} we could explain why a relatively narrow peak
appears in the $\gamma p \to K^+ \Lambda$ reaction on top of a large background and 
it does not show up on top of the broader structure around these energies in the $\gamma p \to K^+ \Sigma^0$ reaction. We could also find an easy interpretation on
why the state does not show up in pion induced reactions, or in reactions with
$\pi N$ or $\eta N$ in the final states, because of the small coupling of the 
$f_0(980)N$ to $\pi \pi$ or of the $a_0(980)N$ to $\eta \pi$. In order to find
extra support for the idea we suggested two experiments. One of them is the
separation of the $S_z=3/2$ and $S_z=1/2$ parts of the $\gamma p \to K^+
\Lambda$ cross section, which would rule out the $3/2^+$ assignment if
there is no cross section in the $S_z=3/2$ channel. The other one is the 
investigation of the cross section close to threshold and the invariant 
mass distributions close to the two kaon mass, which should show enhancements
close to both thresholds, indicating a bound state below the  
$K \bar{K} N $ mass with the two kaons strongly correlated to form the 
$a_0(980)N$ and $f_0(980)N$ states. Both experiments are feasible in existing
laboratories and we hope that the present work encourages their implementation.

\section*{Acknowledgments}  
We would like to thank Juan  Nieves and T. Nakano for useful discussions. This work is 
partly supported by DGICYT contract number
FIS2006-03438. We acknowledge the support of the European Community-Research 
Infrastructure Integrating Activity
``Study of Strongly Interacting Matter" (acronym HadronPhysics2, Grant Agreement
n. 227431) under the Seventh Framework Programme of EU. 
Work supported in part by DFG (SFB/TR 16, ``Subnuclear Structure of Matter''),
and  by the Helmholtz Association through funds provided to the virtual institute
``Spin and strong QCD'' (VH-VI-231).
One of the authors (A.~M.~T.) is supported by a 
FPU grant of the Ministerio de Ciencia y Tecnolog\'ia. 
K.~P.~Khemchandani thanks the support by the 
Funda\c{c}\~{a}o para a Ci\^{e}ncia e a Tecnologia of the Minist\'erio da Ci\^{e}ncia, Tecnologia e 
Ensino Superior of Portugal (SFRH/BPD/40309/2007).

\end{document}